\documentstyle[aps,preprint,epsf]{revtex}

\newcommand{\LSP}{\tilde{\chi}^0}
\newcommand{\postscript}[2]{\setlength{\epsfxsize}{#2\hsize}
  \centerline{\epsfbox{#1}}}

\begin{document}
 
\pagestyle{empty}

\preprint{
\noindent
\hfill
\begin{minipage}[t]{3in}
\begin{flushright}
LBNL--39158 \\
UCB--PTH--96/34 \\
hep-ph/9607453 \\
July 1996
\end{flushright}
\end{minipage}
}

\title{Lepton Flavor Violation at LEP II and Beyond}

\author{Jonathan L.~Feng
\thanks{Research Fellow, Miller Institute for Basic Research in
Science.}
\thanks{This work was supported in part by the Director, Office of
Energy Research, Office of High Energy and Nuclear Physics, Division of
High Energy Physics of the U.S. Department of Energy under Contract
DE-AC03-76SF00098 and in part by the National Science Foundation under
grant PHY-95-14797.}
}
       
\address{
Theoretical Physics Group, LBNL and Department of Physics \\
University of California, Berkeley, California 94720}

\maketitle

\begin{abstract}
If sleptons are produced at LEP II or the Next Linear Collider, lepton
flavor violation can be probed at a level significantly below the
current bounds from rare processes, such as $\mu \to e\gamma$.
Polarizable $e^-$ beams and the $e^-e^-$ mode at the NLC are found to be
powerful options.

\vspace*{.35in}
\centerline{Talk given at the 5th International Workshop on
Supersymmetry}
\centerline{and Unification of Fundamental Interactions (SUSY-96)}
\centerline{University of Maryland, College Park, 29 May -- 1 June 1996}
\end{abstract}


\pagestyle{plain}


At present, two fundamental mysteries in particle physics are the
origins of electroweak symmetry breaking and the fermion mass
matrices. The experimental discovery of superpartners would represent
enormous progress in our understanding of electroweak symmetry breaking,
but would it also allow progress on the flavor problem?  To date, nearly
all experimental studies of supersymmetry have ignored the possibility
of flavor mixings in the sfermion sector.  However, since all
superpartners must be given masses, all supersymmetric theories
necessarily allow for the possibility of new flavor mixings beyond the
standard model.  In addition, there are now many supersymmetric theories
of flavor, which predict a wide variety of superpartner flavor mixings.
In this study, we examine the possibility of measuring these mixings at
LEP II and the Next Linear Collider (NLC).  Rare flavor changing
processes, such as $\mu \to e \gamma$, $\tau \to
\mu \gamma$, $\tau \to e \gamma$, $b \to s \gamma$, and neutral meson
mixing, already provide important constraints on the sfermion flavor
mixings through the virtual effects of superpartners. However, as will
be seen below, once superpartners are discovered, it will be possible to
probe these mixings much more powerfully by directly observing the
change in flavor occurring at the superpartner production and decay
vertices.  This talk is based on work done with N.~Arkani-Hamed,
H.-C.~Cheng, and L.~J.~Hall \cite{paper}.


In the minimal supersymmetric standard model, there are seven new flavor
mixing matrices $W_a, a = u_{L,R}, d_{L,R}, e_{L,R}, \nu_L ,$ where, at
the neutral gaugino vertex of species $a$, the generation $i$ scalar is
converted to the generation $j$ fermion with amplitude ${W_a}_{ij}$.
Clearly, if supersymmetry is correct, it will furnish a large new arena
for studying the problem of flavor.  In this study, we will concentrate
on intergenerational mixing in the slepton sector.  This choice is
motivated by the theoretical prejudice that sleptons are lighter than
squarks and so are presumably more likely to be seen at future
colliders, and also by the absence of standard model lepton flavor
violation, which implies that any flavor violation in the processes we
consider is supersymmetric in origin.  For simplicity, we specialize to
the case of two generation mixing, and in particular,
$\tilde{e}_R$--$\tilde{\mu}_R$ mixing.  We define $\sin \theta_R \equiv
{W_{e_R}}_{12}$, $\Delta m_R^2 \equiv m_{\tilde{e}_R}^2 -
m_{\tilde{\mu}_R}^2$, and $m_R \equiv (m_{\tilde{e}_R} +
m_{\tilde{\mu}_R})/2$.  As in the case of neutrino oscillations, the
power of various experiments is best displayed by plotting their reach
in the $(\sin 2 \theta_R, \Delta m^2_R)$ plane.

The most stringent current bound on $\tilde{e}_R$--$\tilde{\mu}_R$
mixing is from $B(\mu\to e \gamma) < 4.9 \times 10^{-11}$ \cite{LAMPF}.
If $\theta_R$ is comparable to the Cabibbo angle and $\Delta m_R^2 \sim
m_R^2$, the rate for $\mu \to e \gamma$ is typically several orders of
magnitude above the experimental bound. However, for nearly degenerate
$\tilde{e}$ and $\tilde{\mu}$, the bound may be satisfied by a superGIM
cancellation. For large $|\mu|$, two types of diagrams contribute to the
process $\mu\to e \gamma$ \cite{largemu}: one proportional to $\sin 2
\theta_R \Delta m^2_R$, and one proportional to $\sin 2 \theta_R \Delta
m^2_R \tilde{t}$, where $\tilde{t} = -(A + \mu
\tan\beta) / m_R$. These two diagrams may interfere destructively, and
the bound from $\mu \to e \gamma$ disappears for certain
$\tilde{t}$. For $|\mu| \sim m_R, M_1$, an additional diagram weakens
the bounds for large $\tilde{t}$ \cite{smallmu}.

The flavor violating processes that we hope to observe at colliders are
slepton pair production leading to $e^+e^- \to e^{\pm} \mu^{\mp}
\LSP\LSP$ and $e^-e^- \to e^- \mu^- \LSP\LSP$, where for simplicity, we
consider the case where the right-handed sleptons decay directly to the
LSP.  For large $\Delta m_R^2$, the flavor violating cross section is
given simply by multiplying the flavor conserving cross sections by
appropriate factors of the branching ratios $B(\tilde{e} \to \mu \LSP )
= B(\tilde{\mu}\to e \LSP ) =
\frac{1}{2} \sin^2 2\theta_R$.  However, when $\Delta m_R^2 < m_R
\Gamma$, where $\Gamma$ is the slepton decay width, interference effects
become important.  For example, in the $e^-e^-$ case, where sleptons are
produced only through $t$-channel neutralino exchange, the flavor
violating cross section is given by replacing the branching ratios above
with $\frac{1}{2} \sin^2 2\theta_R (\Delta m^2_R)^2 / [(\Delta m^2_R)^2
+ 4m_R^2 \Gamma^2]$, which vanishes in the limit of degenerate sleptons,
as it must.  For $e^+e^-$, the flavor violating cross sections are more
complicated, but have been properly included in this study.  As seen
above, $\Delta m_R^2$ is highly constrained by $\mu\to e\gamma$, and so
the interference effect is important in much of the allowed parameter
space and must be included if one is to assess the ability of future
colliders to detect lepton flavor violation.


Having discussed the flavor violating cross sections, we now examine the
possibility of detecting such flavor violating signals at future
colliders.  We first consider the sensitivity of the LEP II $e^+e^-$
collider, with a center of mass energy $\sqrt{s} = 190 \, \rm{ GeV}$ and
an integrated luminosity of roughly $500 \, \rm{ pb}^{-1}$.  To discuss
the flavor violation discovery potential of LEP II, we must first choose
some representative values for the various SUSY parameters.  Sleptons
with mass below 85--90 GeV are expected to be discovered at LEP II. We
therefore consider the case where $m_{\tilde{e}_R} \approx
m_{\tilde{\mu}_R} \approx 80 \, \rm{ GeV}$.  The LSP must be lighter
than this, and we assume that the LSP is roughly Bino-like with mass
$M_1 = 50$ GeV.  These mass choices are representative of the parameter
space available to LEP II, and lead to conclusions which are neither
pessimistic nor optimistic.  Implications of deviations from these
choices are discussed in Ref.~\cite{paper}.

\begin{figure}[htb]
\vspace{1in}
\postscript{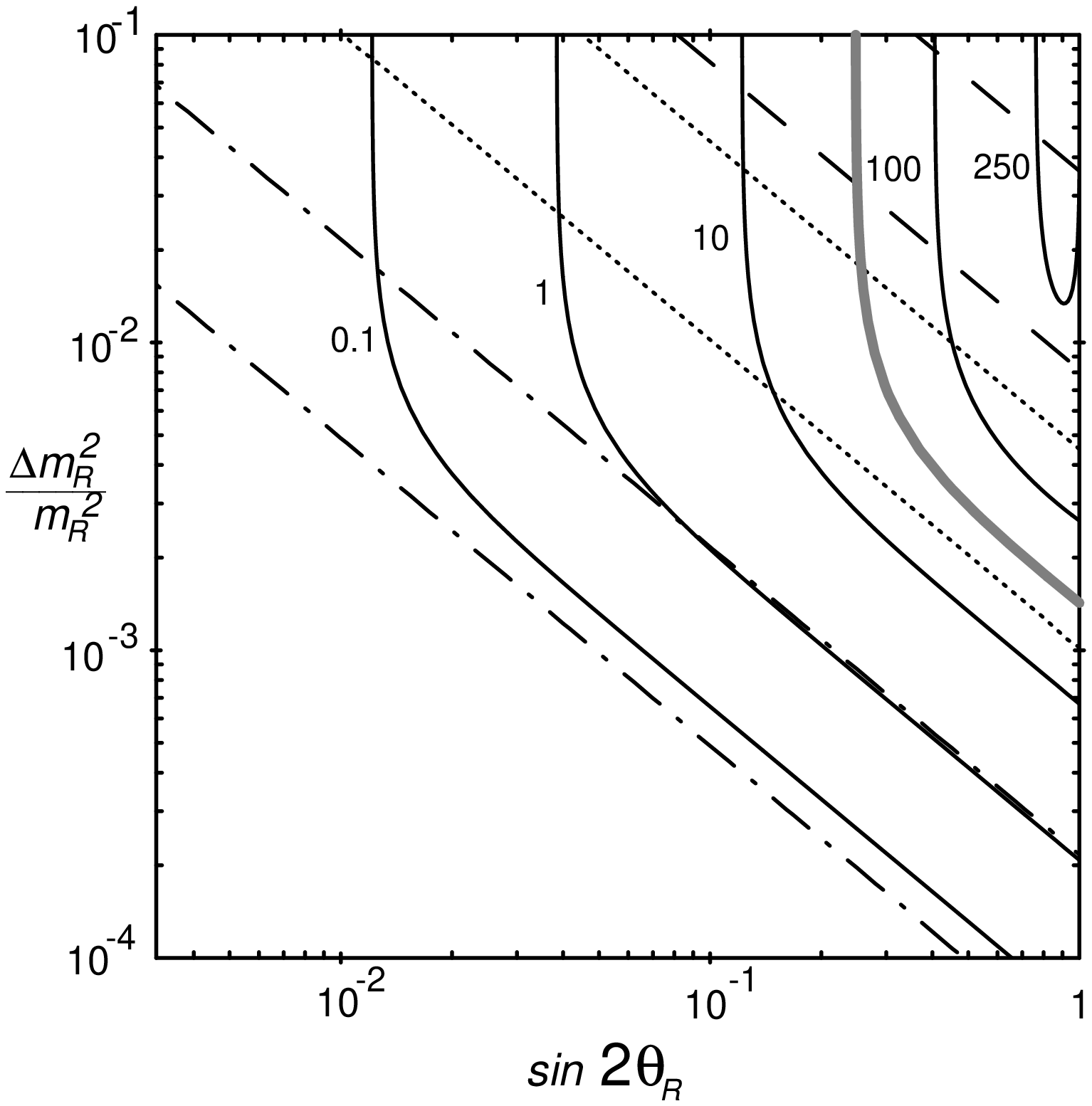}{0.8}
\vspace*{.3in}
\caption{Contours of constant $\sigma ( e^+e^- \to e^{\pm} \mu^{\mp} 
\tilde{\chi}^0\tilde{\chi}^0 )$ (solid) in fb for LEP II, with 
$\protect\sqrt{s} = 190 \, \rm{ GeV}$, $m_{\tilde{e}_R}, m_{\tilde{\mu}_R}
\approx 80 \, \rm{ GeV}$, and $M_1 = 50 \, \rm{ GeV}$.  The thick gray 
contour represents the (optimal) experimental reach for one year's
integrated luminosity.  Constant contours of $B(\mu \to
e\gamma)=4.9\times 10^{-11}$ and $2.5\times 10^{-12}$ are also plotted
for $\tilde{t} \equiv -(A + \mu \tan\beta)/\bar{m}_R = 0$ (dotted), 2
(dashed), and 50 (dot-dashed) and degenerate left-handed sleptons with
mass 120 GeV.}
\label{fig:LEPII}
\end{figure}

At LEP II energies, the dominant standard model background to the
$e^{\pm} \mu^{\mp}\LSP \LSP$ final state is $W$ pair production, where
both $W$ bosons decay to $e$ or $\mu$, either directly or through $\tau$
leptons.  Including branching ratios, this cross section is 680 fb.  The
$W^+W^-$ background may be reduced with cuts, as has been discussed in a
number of studies \cite{LEPII}.  Depending on the LSP mass, the cuts may
be optimized to reduce the background to $\sim$10--100 fb, while
retaining 40\% -- 60\% of the signal.  Given an integrated luminosity of
$500 \, \rm{ pb}^{-1}$, the required cross section for a $5\sigma$
effect is $\sim$40--185 fb.  The flavor violating cross section is
plotted in Fig.~\ref{fig:LEPII}, along with the constraint from
$B(\mu\to e\gamma)$ for various values of $\tilde{t}$. The cross section
contours extend to $\sin\theta_{R}\sim 0.15$, and, for low values of
$\tilde{t}$, probe new regions of parameter space.  Surprisingly, we
find that LEP II, which produces merely a few hundred sleptons a year,
may be able to detect lepton flavor violation.

If sleptons are not produced at LEP II, they may be discovered at the
NLC.  There are many options at the NLC, as both highly polarized $e^-$
beams and $e^+e^-$ and $e^-e^-$ modes may be available.  We will assume
the NLC design energy $\sqrt{s} = 500 \, \rm{ GeV}$ and luminosity $50
\, (20)\, \rm{ fb}^{-1}/\rm{yr}$ in $e^+e^-$ ($e^-e^-$) mode. For the NLC,
we consider right-handed slepton masses $m_{\tilde{e}_R} ,\,
m_{\tilde{\mu}_R} \approx 200 \, \rm{ GeV}$, and $M_1= 100 \, \rm{
GeV}$.

We consider first the $e^+e^-$ mode. At NLC energies, $W^+W^-$, $e^{\pm}
\nu W^{\mp}$, and $(e^+e^-) W^+ W^-$ all are significant backgrounds.
Nevertheless, efficient cuts \cite{BV} and a right-polarized $e^-$ beam
effectively isolate the flavor violating signal.  Given a year's running
at design luminosity, the required $5\sigma$ signal is 3.8 (3.6) fb for
90\% (95\%) right-handed beam polarization.  The NLC in $e^+e^-$ mode is
a powerful probe of the flavor violating parameter space, extending to
$\sin\theta_{R} = 0.06$ and probing parameter space for which $B(\mu \to
e\gamma) = 10^{-14}$ ($10^{-11}$) for $\tilde{t} = 2$ (50).

\begin{figure}[htb]
\vspace*{1in}
\postscript{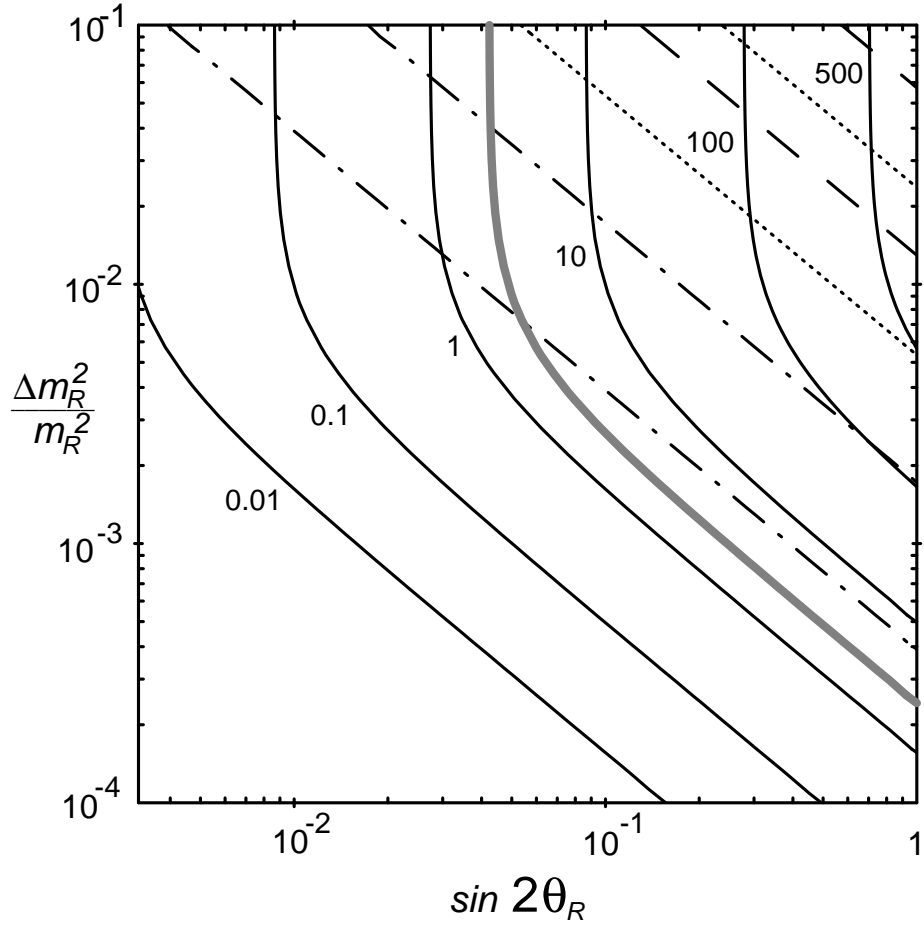}{0.8}
\vspace*{.3in}
\caption{Same as in Fig.~\protect\ref{fig:LEPII}, but for the NLC in
$e^-_R e^-_R$ mode, with $\protect\sqrt{s} = 500 \, \rm{ GeV}$,
$m_{\tilde{e}_R}, m_{\tilde{\mu}_R} \approx 200 \, \rm{ GeV}$, $M_1 = 100
\, \rm{ GeV}$, and left-handed sleptons degenerate at 350 GeV.}
\label{fig:NLC-}
\end{figure}

An intriguing feature of the NLC is its ability to run in $e^-e^-$ mode.
Slepton pair production is allowed, as SUSY theories naturally provide
Majorana particles, the neutralinos, which violate fermion
number. However, many troublesome backgrounds, for example, $W$ pair
production, are completely eliminated.  In addition, this option allows
one to polarize both beams.  For RR beam polarization, there are
essentially no backgrounds; the dominant background is $e^-\nu
W^-$\cite{CC}, arising from imperfect beam polarization.  With 90\%
(95\%) right-polarized beams and without additional cuts, the required
$5\sigma$ signal for one year's luminosity is 3.9 (2.5) fb. The reach in
parameter space is shown in Fig.~\ref{fig:NLC-}.  This proves to be the
most sensitive mode considered so far, probing mixing angles with
$\sin\theta_{R} = 0.02$ and parameter space for which $B(\mu \to
e\gamma) = 10^{-15}$ ($ 10^{-12}$) for $\tilde{t} = 2$ (50), far below
the current bounds.  We find that, if sleptons are kinematically
accessible at the NLC, the $e\mu$ signal will provide either stringent
upper bounds on slepton mixing or the exciting discovery of
supersymmetric lepton flavor violation.

\end{document}